\documentclass{iaus}
\usepackage{graphicx}

\title[Lopsidedness and sloshing in centres of mergers]
{Lopsidedness and Sloshing in Centres of Advanced Mergers of Galaxies}

\author[Jog \& Maybhate]
{Chanda J. Jog$^1$ \& Aparna Maybhate$^2$}

\affiliation{$^1$Department of Physics, Indian Institute of Science, Bangalore 560012, India \break 
$^2$Space Telescope Science Institute, Baltimore, MD 21218, U.S.A. \break email: 
cjjog@physics.iisc.ernet.in, maybhate@stsci.edu}

\pubyear{2007}
\volume{245}  
\pagerange{-----}
\date{?? and in revised form ??}

\jname{Proceedings Title IAU Symposium}
\editors{M. Bureau, E. Athanassoula, \& B. Barbuy, eds.}
\begin{document}

\maketitle

\begin{abstract}
We measure the non-axisymmetry in the luminosity distribution in the central few kpc of 
a sample of advanced mergers of 
galaxies, by analyzing their 2MASS images. All mergers show a high central asymmetry:
the centres of isophotes show a
striking sloshing pattern with a spatial variation of upto 30\% within the central 1 kpc;
and the Fourier amplitude for lopsidedness ($m=1$) shows high values upto 0.2 within the central 5 kpc. 
The central asymmetry is estimated to be long-lived, lasting for $\sim$ a few Gyr
or $\sim$ 100 local dynamical timescales. This will significantly affect the
dynamical evolution of this region, by helping fuel  the
central active galactic nucleus, and also by causing the secular growth of the bulge driven by lopsidedness.
\end{abstract}

\firstsection 
\section{Introduction}
Interactions and mergers of galaxies are known to be common and these significantly affect
their dynamics and evolution.
The outer regions of merger remnants covering a radial range of $\sim $ a few kpc to a few $\times$ 10 kpc have 
been well-studied. These can be fit by an elliptical-like r$^{1/4}$ profile (class I), or a disc-like outer exponential profile (class II), or a no-fit profile (class III)
(Chitre \& Jog 2002). The first two can be explained as arising due to equal-mass mergers (e.g., Barnes 1992)
or unequal-mass mergers (Bournaud, Combes \& Jog 2004) respectively, while the third case corresponds to younger remnants. 
However, the central regions of a few kpc in mergers have not been studied in detail so far - which motivated our work (Jog \& Maybhate 2006) presented here, where we study
a sample of advanced mergers which show signs of recent interaction such as tidal tails or loops but have a single nucleus.

\section {Analysis \& Results }
Elliptical isophotes were fit to the K$_s$-band images from 2MASS 
while allowing the centre, ellipticity and the position angle to vary to get the best fit. Figure 1 (top panel) shows the result for Arp 163 (a class III galaxy) -
the isophotes are not concentric, instead the centres (X$_0$,Y$_0$) of consecutive isophotes 
show a wandering or sloshing pattern with a $\sim 30 \%$ variation within the central 1 kpc, indicating an unrelaxed central region. Further, we measure  the lopsidedness of the distribution by Fourier-
analyzing the galaxy image
 w.r.t. a constant centre. The amplitude A$_1$ and the phase p1 for $m=1$ were plotted versus radius- as shown for Arp 163 in the lower panel of Figure 1, which shows high central lopsidedness $\sim 0.15$ over the central 5 kpc region.

All the sample galaxies show strong sloshing and lopsidedness in the central regions. The
asymmetry does not depend significantly on the masses of the progenitor galaxies, being similar for class I and II cases, but it is higher for younger mergers (class III).
The corresponding values
are smaller by a factor of few for a control sample of non-merger galaxies,
confirming the merger origin of the high central asymmetry in our sample.

\begin{figure}
\includegraphics[height=4.3in,width=4.3in]{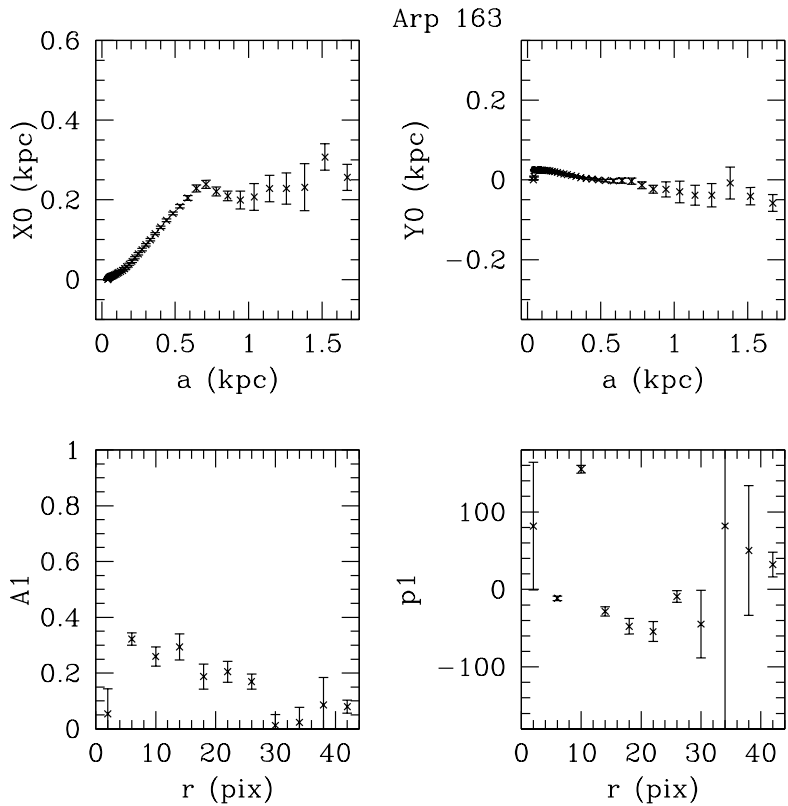} 
\caption{Arp 163: The centres of the isophotes vs. the semi-major axis (top panel):  the centres show a sloshing pattern indicating an unrelaxed region;  the plot of Fourier amplitude $A_1$ and the phase $p_1$ vs. radius (lower panel)    shows a high lopsided amplitude  
and a fluctuating phase.
}
\end{figure}

The ages of remnants are deduced to be $\sim$ 1-2 Gyr as seen from the merger remnants with similar outer disturbed features in the N-body simulations (Bournaud, Combes \& Jog 2004).
Thus the central asymmetry lasts for over $\sim 100$ local dynamical timescales, and  will have important consequences for the evolution of the central region. First, it can help fuel the central active galactic nucleus, 
and second,  it can lead to the secular growth of the bulge via 
the lopsided distribution. These need to be studied in detail theoretically. Since this predicted evolution is due to the central asymmetry that is merger-driven, it could be important  in the hierarchical evolution of galaxies.

\begin{acknowledgments}
CJ would like to acknowledge the support from the IAU to attend this meeting.
\end{acknowledgments}


\begin{thebibliography}{}

\bibitem[Barnes 1992]{Barnes02}
       {Barnes, J.} 1992, \textit{ApJ}, 393, 484

\bibitem[Bournaud, Combes \& Jog (1994)]{Bournaud04}
     {Bournaud, F., Combes, F., \& Jog, C.J.} 2004,
     \textit{A \& A}, 418, L27


\bibitem[Chitre \& Jog (2002)]{Chitre02}
         {Chitre, A. \& Jog, C.J.} 2002, \textit{A \& A}, 388, 407




\bibitem[Jog \& Maybhate]{Jog06}
     {Jog, C.J. \& Maybhate, A.} 2006,
     \textit{MNRAS}, 370, 891



\end{thebibliography}
\end{document}